\documentclass[prc,twocolumn,showpacs,floatfix,nofootinbib,preprintnumbers,superscriptaddress,amsmath,amssymb]{revtex4}

\usepackage{CJKutf8}
\usepackage{graphicx}
\usepackage{epsfig}
\usepackage{amsmath,amssymb}
\usepackage{bm}
\usepackage{color}
\usepackage{float}
\usepackage{dcolumn}
\usepackage{multirow}
\usepackage{mathrsfs}
\usepackage[arrowdel]{physics}
\usepackage[OT1]{fontenc}
\usepackage{outline}
\usepackage{tikz,tikz-3dplot}
\usetikzlibrary{positioning}

\DeclareMathAlphabet{\mathcal}{OMS}{cmsy}{m}{n}

\newcommand{\HFODD}{\textsc{hfodd}}

\begin{document}
\begin{CJK*}{UTF8}{gbsn}

\title{Precision of finite-difference representation in 3D coordinate-space Hartree-Fock-Bogoliubov calculations}

\author{Yue Shi (石跃)}
\email{yueshi@hit.edu.cn}
\affiliation{Department of Physics, Harbin Institute of Technology, Harbin 150001, People's Republic of China}

\begin{abstract}
\begin{description}
\item[Background] The precision of nuclear Hartree-Fock-Bogoliubov
  (HFB) calculations in coordinate space is limited by box
  discretization schemes. In particular, for finite-difference (FD)
  discretization method, the resolution and box size determines the
  calculation error.

\item[Purpose] The current work plans to study the accuracy due to FD
  approximation to the 3D nuclear HFB problem.

\item[Methods] By (1) taking the wave functions solved in harmonic
  oscillator (HO) basis, (2) representing the HFB problem in
  coordinate space using FD method, the current work carefully
  evaluates the error due to box discretization by examining the
  deviation of the resulted HFB matrix, the total energies in the
  coordinate space, from those calculated with HO method, the latter
  of which is free from numerical error within its model
  configuration. To estimate how the error (given by the box
  discretization schemes suggested above) accumulates with
  self-consistent iterations, self-consistent HF and HFB calculations
  (with two-basis method) has been carried out for doubly magic
  nuclei, $^{40}$Ca, $^{132}$Sn, and $^{110}$Mo. The resulted total
  energies are compared with those of HO basis, and 3D coordinate
  space calculations in literatures.

\item[Results] The analysis shows that, for grid spacing $\le$0.6\,fm,
  the off-diagonal elements of the resulted HFB matrix elements (M.E.)
  are extremely small ($<$1\,keV). The resulted quasi-particle (q.p.)
  spectra differ from those of HO calculations by a few
  keV. Self-consistent HF and HFB calculations within the current FD
  method with the above box discretizatioin schemes give results
  similar to those calculations of existing HO basis, and coordinate
  space method. For the HFB calculations, it is demonstrated that the
  FD method and the HO method predicted different single-particle
  spectra, which makes an exact comparison difficult. This makes the
  analysis of precision at a certain iteration useful.

\item[Conclusions] With the described FD approximation to the
  differential operators, together with the way various densities, and
  Hamiltonian are constructed in the current work, it can be concluded
  that for box grid spacing $\le$0.7\,fm, and large enough box size
  for the studied system, the accuracy of each calculated energy
  contribution is in the order of a few tens of keV. With this
  discretization scheme, the number of the calculated M.E. differs
  from that of HO calculation by $\le$10\% of the total number of the
  HFB M.E. For the single-particle or q.p. energies, the accuracy is
  in the order of a few keV. The above conclusions have been verified
  by performing self-consistent HF, and HFB calculations.

\end{description}
\end{abstract}


\maketitle
\end{CJK*}

\section{Introduction}
\label{sec1}

Nuclear density functional theory (DFT) or nuclear self-consistent
Hartree-Fock (HF) theory~\cite{bend03} has been successful in
describing ground-state (g.s.)  properties in nuclei throughout the
nuclear chart.
After incorporating pairing interaction in the frameworks of
Bardeen-Cooper-Schriffer (BCS) or Hartree-Fock-Bogoliubov
(HFB)~\cite{doba84}, the resulted HF+BCS or HFB theories have become
standard models for the description of low-energy phenomena in nuclear
physics. In addition, the obtained static solutions provide useful
starting points for beyond mean-field calculations~\cite{bend03,
  ring80}. Indeed, static solutions of the g.s. provide starting
point for time-dependent analysis which is essential for studying
nuclear reaction and fissioning processes~\cite{bonche76, maruhn14}.

Various mean-field and HFB methods require a fast determination of the solution
of Schr\"odinger equation. In theoretical nuclear physics, two main types of
methods exist to solve the non-linear problem.  On the one hand, there are
methods based on harmonic oscillator (HO) basis expansion~\cite{doba95,
doba97a, doba97b, doba00, doba04, doba05, doba09, schunck12, schunck17, stoi05,
stoi13}, or, more generally, on the use of analytic basis~\cite{bulg13}. On the
other hand, there are methods that express the w.fs. and operators in
coordinate space directly, or using interpolatory techniques~\cite{bonche85,
gall94, tera96, yama01, teran03, ober03, taji04, pei08, pei14, oba09, zhang11,
zhang12, ryss15a, ryss15b, jin17}.

Methods based on HO basis possess the following main advantages. (1)
Within the Skyrme energy density functional (EDF) theory, the nuclear
mean-field Hamiltonian can be evaluated {\it exactly} without
approximation in HO basis~\cite{doba97a}; and (2) by limiting the
number of HO, the basis expansion methods provide a natural truncation
for the problem, which leads to a Hamiltonian matrix with reasonably
small size, providing acceptable and tunable precision. Methods using
HO basis suffer from strong damping at large space distance from the
center of nucleus, so that the precision of these methods may not be
guaranteed for systems with large density extension.

Coordinate space methods are complementary when it comes to their
numerical properties. Firstly, the representation of w.fs. in
coordinate space allows for more extensive w.fs. in real
space. Secondly, these methods provide intuitive, and hence,
relatively straightforward codes, facilitating further updates.
Thirdly, evaluation of quantities on grids could be performed
separately, providing excellent parallel scalability, which is
particularly suitable for modern super-computers. However, there are
two main sources of error related to methods in coordinate space,
namely, box size, and grid spacing.

Recently, there has been new efforts in basis optimization for 3D
HF(B) calculations, especially using coordinate space~\cite{jin17,
  bulg13, ryss15a, ryss15b,pei14,schu15,schu16,afib18}. Taking
advantage of the increasingly larger computing resources involves
compromise between the degree of complexity of method used, and the
precision that can be accepted for applications at hand. Hence, it is
necessary to know the precision of the nuclear HF(B) calculation
induced by limited resolution and box size.

Conventionally, one estimates the errors of a self-consistent HF(B)
calculation by comparing its calculated results with a presumably
more accurate calculation with a lower dimensioned method, allowing
for much better resolution in coordinate space~\cite{doba84, benn05},
the latter being considered close to an exact solution.

However, except for extreme cases \cite{schunck17, bulg13} where the
basis can be made exactly the same, HF(B) calculations of different
dimensions can hardly be made identical. This is due to different
truncation schemes used for the continuum spectrum, which results in
different (quasi-)particle levels retained in respective Fock
spaces. This makes strict benchmark calculations extremely difficult
to devise, particularly for applications using Cartesian coordinates.

A previous study \cite{blum92} considered the precision of various
techniques including the finite difference (FD) method. In their
study, a Woods-Saxon form of mean-field was used to simulate that
resulting from an EDF calculation. In the current study, I plan to
look into the errors due to box discretization for EDF. In
Sec.~\ref{model}, the nuclear DFT is briefly reviewed. The current
w.fs., the definition of derivative, and Laplacian operators, and the
strategy to quantify precisions will be given at the end of the
section. The results of error analysis about FD method will be shown
in Sec.~\ref{results}. Then, the self-consistent calculations of
Skyrme HF, and HFB using two-basis method will be presented in
Sec.~\ref{examples}, before the summary and perspectives, which are
shown in Sec.~\ref{conclusions}.

\section{The model}
\label{model}

The current work aims at studying the precisions associated with grid
spacing and box size in 3D coordinate space HFB calculations. Before
presenting the strategy followed to evaluate its precision, in this
section, I will recall briefly the formalism of the nuclear Skyrme HFB
problem.

\subsection{Quasi-particle states}

The q.p. w.fs. are primary inputs for the evaluation of local
densities etc. In practice, one uses well-educated initial results
from Nilsson, Woods-Saxon, or codes of a lower dimension. In the
present work, I plan to estimate the error due to limited resolution
of FD procedure by comparing results with those from 3D HO basis.

The w.fs. in coordinate space [Eq.~(10) in Ref.~\cite{doba04}] are
implemented using the expansion coefficients ($A$'s and $B$'s) using
HO basis resulting from a converged {\HFODD} (v2.49t)~\cite{doba09}
result. Specifically, the upper and lower components of w.fs. in
coordinates space read
\begin{eqnarray}
\label{wf}
\varphi^{(1)}_{\alpha,s=\pm i}(\vec{r}\sigma) &=& 
          2 \sigma \sum_{\vec{n}} \psi^*_{\vec{n},s=\pm i}(\vec{r},-\sigma)A^*_{\pm,\vec{n}\alpha}, \nonumber \\
\varphi^{(2)}_{\alpha,s=\pm i}(\vec{r}\sigma) &=& 
          \sum_{\vec{n}} \psi^*_{\vec{n},s=\pm i}(\vec{r},\sigma)B^*_{\pm,\vec{n}\alpha},
\end{eqnarray}
where $\psi_{\vec{n},s=\pm i} (\vec{r}\sigma)$ are the HO w.fs. [Eq.
(78) in Ref.~\cite{doba97a}] in space coordinates [Eq. (76) in
Ref.~\cite{doba97a}] and $\vec{n}=(n_x, n_y, n_z)$ are the HO quanta
numbers in three Cartesian directions. The index $\alpha=1,...,M/2$
numbers eigenstates of HFB equations. The number of q.p. states
included in the HFB equations, $M$, is defined by a cut-off energy
$E_{\rm cut}$.

\subsection{Local densities}

Local densities for particle, pairing, kinetic are obtained from
q.p. w.fs.,
\begin{eqnarray}
\rho(\vec{r}) &=& \sum_{\alpha,s=\pm i}\sum_{\sigma=\pm\frac{1}{2}} 
                  \qty|\varphi^{(2)}_{\alpha,s} (\vec{r}\sigma)| ^2,\\
		  \label{tau}
\tau(\vec{r}) &=& \sum_{\alpha,s=\pm i}\sum_{\sigma=\pm\frac{1}{2}} 
                  \qty|\vec{\nabla} \varphi^{(2)}_{\alpha,s} (\vec{r}\sigma)| ^2,\\
\tilde{\rho}(\vec{r}) &=& -2\sum_{\alpha}\sum_{\sigma=\pm\frac{1}{2}}  
\varphi^{(2)}_{\alpha,s=+i} (\vec{r}\sigma) \varphi^{(1)*}_{\alpha,s=-i} (\vec{r}\sigma).
\end{eqnarray}

For the present applications, spin-orbit density does not appear
explicitly. The divergence of spin-orbit density reads
\begin{eqnarray}
\vec{\nabla}\cdot\vec{\mathrm{J}}=-i\sum_{\alpha,\sigma,s=\pm i} \qty(\vec{\nabla}\varphi^{(2)*}_{\alpha,s} (\vec{r}\sigma))
\cdot \qty(\vec{\nabla} \cross \vec{\sigma})\varphi^{(2)}_{\alpha,s} (\vec{r}\sigma).
\end{eqnarray}

To this point the index ($q$) denoting neutron ($n$) and proton ($p$)
has been ignored for simplicity.

\subsection{The energy-density functional}

For Skyrme-HFB calculation, the total energy $E$ of a nucleus is the
sum of kinetic, Skyrme, pairing, and Coulomb terms:
\begin{eqnarray}
\label{Etotal}
E &=& E_{\rm Kin+c.m.} + E_{\rm Skyrme} + E_{\rm pair} + E_{\rm Coul} \nonumber \\
  &=& \int \dd[3]{\vec{r}} \big[\mathcal{K}(\vec{r}) + \mathcal{E}_{\rm Skyrme}(\vec{r}) \nonumber \\
      & &   + \mathcal{E}_{\rm pair}(\vec{r}) + \mathcal{E}_{\rm Coul}(\vec{r})\big].
\end{eqnarray}
The derivation of the total energy has been presented in detail in
Refs.~\cite{vaut72, engel75} for the HF case, and in
Ref.~\cite{doba84} for the HFB case.

In Eq.~(\ref{Etotal}), the kinetic energies of both neutron and proton
are given by
\begin{eqnarray}
\label{kinetic}
\mathcal{K} = \frac{\hbar^2}{2m} \tau \qty(1-\frac{1}{A}),
\end{eqnarray}
where the factor in between parentheses takes into account the
one-body part of center-of-mass correction~\cite{benn05}. As the
current work only considers even-even nuclei, the Skyrme part of the
EDF is time even. The energy density functional reads
\begin{eqnarray}
\label{skyrme}
\mathcal{E}_{\rm Skyrme} &=& \frac{b_0}{2} \rho^2 - \frac{b'_0}{2}\sum_q \rho^2_q 
                           + \frac{b_3}{3}\rho^{\alpha+2} - \frac{b'_3}{3}\rho^{\alpha}\sum_q \rho^2_q \nonumber \\
               && + b_1\qty(\rho\tau-j^2) - b'_1\sum_q\qty(\rho_q\tau_q-j^2_q) \nonumber \\
               && - \frac{b_2}{2} \rho\nabla^2\rho \nonumber + \frac{b'_2}{2}\sum_q\rho_q\nabla^2\rho_q \nonumber \\
               && - b_4\rho\vec{\nabla}\cdot\vec{\mathrm{J}} - b'_4\sum_q\rho_q\qty(\vec{\nabla}\cdot\vec{\mathrm{J}}_q).
\end{eqnarray}
The index $q$ denotes neutron and proton. The densities without index
indicate the sum of neutron and proton densities. The time-odd current
densities $(\vec{j}, \vec{j}_q)$ are identically zero for the current
time-independent study.

The pairing density is~\cite{teran03}
\begin{eqnarray}
\mathcal{E}_{\rm pair} = \sum_q \frac{V^q_0}{4}\tilde{\rho}^2_q(\vec{r})f(\vec{r}).
\end{eqnarray}
Volume pairing is used in the current study, so that $f(\vec{r})=1$.

\subsection{HFB mean-fields}

Varying the total energy Eq.~(\ref{Etotal}) with respect to $\rho$ and
$\tilde{\rho}$ one obtains the HFB equation~\cite{ring80,doba84}
\begin{equation}
\label{HFB}
\comm{\mathscr{W}}{\mathscr{R}}=0,
\end{equation}
where
\begin{equation}
\label{W}
\mathscr{W}=
\mqty(h-\lambda & \tilde{h} \\ \tilde{h} & -h+\lambda),
\end{equation}
and
\begin{equation}
\label{densities}
\mathscr{R}=
\mqty(\rho & \tilde{\rho} \\ \tilde{\rho} & \delta(\vec{r}-\vec{r'})\delta_{\sigma\sigma'}-\rho).
\end{equation}
In Eq.~(\ref{W}), the Lagrange multiplier $\lambda$ guarantees
$\int\dd[3]\vec{r}\sum_{\sigma}\rho(\vec{r}\sigma,\vec{r}\sigma)$ to
be equal to the proton or neutron number.

The mean-fields and densities appearing in
Eqs.~(\ref{HFB},\ref{W},\ref{densities}) are not necessarily
local~\cite{doba84,perl04}. But due to the zero-range of Skyrme force
used in the current work, the densities in Eq.~(\ref{densities}) can
be expressed using only local particle and pairing
densities~\cite{vaut72,doba84}
\begin{eqnarray}
\rho(\vec{r}\sigma,\vec{r}'\sigma') &=& \frac{1}{2} \rho(\vec{r})\delta(\vec{r}-\vec{r}')\delta_{\sigma\sigma'}, \\
\label{pdensity}
\tilde{\rho}(\vec{r}\sigma,\vec{r}'\sigma') &=& \frac{1}{2} \tilde{\rho}(\vec{r})\delta(\vec{r}-\vec{r}')\delta_{\sigma\sigma'},
\end{eqnarray}
where the time-odd parts have been ignored as they identically vanish
in a static study. In Eq.~(\ref{W}) the mean-field Hamiltonian ($q$
distinguishing $n$, and $p$) reads
\begin{eqnarray}
\label{mean-field}
h_q\qty(\vec{r}\sigma,\vec{r}'\sigma') =&& 
        \qty[- \vec{\nabla} \cdot \frac{\hbar^2}{2m^*_q}\vec{\nabla} + U_q]
                                      \delta(\vec{r}-\vec{r}')\delta_{\sigma\sigma'} \nonumber \\
      &&- \qty[i \vec{\mathrm{B}}_q\qty(\vec{\nabla}\cross \vec{\sigma})]_{\sigma\sigma'} \delta(\vec{r}-\vec{r}'),
\end{eqnarray}
where the effective mass is defined by
\begin{equation}
\frac{\hbar^2}{2m^*_q} = \frac{\hbar^2}{2m_q} + b_1 \rho - b'_1 \rho_q,
\end{equation}
and the effective spin density
\begin{equation}
\vec{\mathrm{B}}_q = b_4 \vec{\nabla}\rho + b'_4 \vec{\nabla}\rho_q.
\end{equation}
The nuclear potential due to Skyrme force reads 
\begin{eqnarray}
U_q &=& b_0 \rho - b'_0 \rho_q + b_1 \tau - b'_1 \tau_q \nonumber \\
    & & + \frac{b_3}{3}\qty(\alpha+2)\rho^{\alpha+1}
        - \frac{b'_3}{3}\qty[\alpha\rho^{\alpha-1} \sum_q \qty(\rho^2_q+2\rho^{\alpha} \rho_q)] \nonumber \\
    & & -b_4 \vec{\nabla} \cdot \vec{\mathrm{J}} - b'_4\vec{\nabla}\cdot\vec{\mathrm{J}}_q \nonumber \\
    & & -b_2\nabla^2\rho + b'_2\nabla^2\rho_q.
\end{eqnarray}

Note that the Hamiltonian in Eq.~(\ref{mean-field}) is a $2\cross2$
matrix in $\sigma$-space. The pairing mean-field, which is diagonal in
$\sigma$-space, reads
\begin{equation}
\label{phamiltonian}
\tilde{h}_q\qty(\vec{r}\sigma,\vec{r}'\sigma')  = 
            \frac{1}{2}V^q_0\tilde{\rho}_q(\vec{r}) f(\vec{r})\delta(\vec{r}-\vec{r}')\delta_{\sigma\sigma'}.
\end{equation}

\subsection{Coulomb potential and energy}

The protons potential includes Coulomb contributions. Its direct part
is obtained by solving the electrostatic Poisson equation
\begin{equation}
\Delta U_{\rm Coul}^{(\rm Dir.)}(\vec{r}) = - 4 \pi e^2 \rho_p(\vec{r}).
\end{equation}
The boundary conditions need to be imposed according to
\begin{equation}
U(\vec{r})=\frac{e^2 Z}{r} + e^2\frac{\expval{\hat{Q}_{20}}Y_{20}(\vec{r}) + \expval{\hat{Q}_{22}}\mathcal{R}Y_{22}(\vec{r})}{r^3},
\end{equation}
where the multiple moments $\hat{Q}_{20}$ and $\hat{Q}_{22}$ are
defined by using the spherical harmonics,
$\hat{Q}_{\lambda\mu}=r^{\lambda}Y_{\lambda\mu}$, specifically,
\begin{eqnarray}
\label{Q20}
\hat{Q}_{20}&=&\sqrt{\frac{5}{16\pi}}(2z^2-x^2-y^2), \\
\label{Q22}
\hat{Q}_{22}&=&\sqrt{\frac{15}{32\pi}}(x^2-y^2).
\end{eqnarray}
For details, see Ref.~\cite{ryss15a}.

The exchange part of the Coulomb potential is approximated by Slater
approximation. Its contribution to the Coulomb potential is
\begin{equation}
U_{\rm Coul}^{(\rm Exc.)} = -e^2\qty(\frac{3}{\pi})^{1/3} \qty[\rho_p(\vec{r})]^{1/3}.
\end{equation}

The contributions of direct and exchange Coulomb energies are,
\begin{equation}
E_{\rm Coul} = E_{\rm Coul}^{\rm (Dir.)} + E_{\rm Coul}^{\rm (Exc.)},
\end{equation}
where
\begin{equation}
E_{\rm Coul}^{\rm (Dir.)} = \frac{1}{2}\int {\rm d}\vec{r} U_{\rm Coul}^{\rm (Dir.)}\rho_p(\vec{r}),
\end{equation}
and
\begin{equation}
E_{\rm Coul}^{\rm (Exc.)} = -\frac{3e^2}{4} \qty(\frac{3}{\pi})^{1/3} \int {\rm d}\vec{r} \rho_p^{4/3}(\vec{r}).
\end{equation}

\subsection{Finite-difference discretization of q.p. states and operators}

The current work adopts FD discretization in coordinate space. Though
FD are known to be inferior in terms of precision compared to Fourier
transformations or spline techniques~\cite{blum92}, it provides
reasonable balance between cost and precision. In computational
intense simulations such as TDHF, and 3D HFB, FD is still widely
used~\cite{maru14,godd15}. It is the purpose of the present work to
examine (1) that to what extend a simple, and efficient FD method
could approach a self-consistent HF(B) problem, and (2) its accuracy.

\begin{figure}
\centering
\includegraphics[width=0.85\columnwidth]{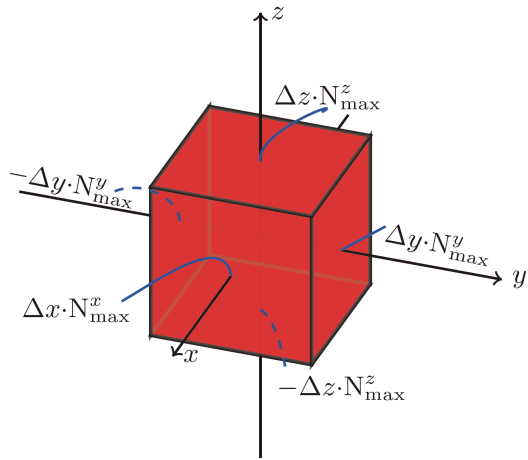}
\caption{Schematic figure depicting grid spacings, $\Delta x$, $\Delta y$, and $\Delta z$, and
box size (through $N^x_{\rm max}$, $N^y_{\rm max}$, and $N^z_{\rm max}$) in coordinate space method.}
\label{figure1}
\end{figure}

Figure~\ref{figure1} illustrates how the cubic and uniformly placed
grid points are defined by grid spacings, $\Delta x,y,z$ (in fm), and
$N^{x,y,z}_{\rm max}$ which numerate the outer-most grid point in each
axis. The box size is thus $\qty(2\Delta x \cdot N^x_{\rm max}) \cross
\qty(2\Delta y \cdot N^y_{\rm max}) \cross \qty(2\Delta z \cdot
N^z_{\rm max})$ (fm$^3$). No symmetry has been enforced in the current
work.

The derivative, and Laplacian operations are evaluated on grids using
seven-, and nine-point formula, respectively~\cite{abra65}. These
choices have been proven to be stable, and reasonably accurate within
FD method from a few groups in different applications~\cite{bonche85,
  blum92, tera96, bonche05, godd15}. The current work examines
precisions of the same setup (seven-, and nine-point formula) by
comparing the results with {\HFODD} while varying the discretization
schemes, at a fixed iteration. Note that, even though the operators
are constructed the same way, how the Hamiltonian is calculated
numerically may give rise to errors. For example, two different
methods to construct kinetic density (\ref{tau}) result in the
kinetic energies (\ref{kinetic}) being different by a few MeV, as
shown table~\ref{table1}.

The integrations are performed using the trapezoidal
rule~\cite{wong97}. Specifically, the summed value at the center of
each cell is evaluated by averaging values at the eight surrounding
corners~\cite{wiki_interpolation}. This procedure differs from other
existing FD implementations, where rectangular rules are
used~\cite{bonche85,tera96,bonche05,maru14,godd15}. Representing the
problem {\it on} the grid points instead of the center allows for more
elaborate interpolatory and extrapolatory methods~\cite{wong97}.

The error associated with this non-conventional way of integration can
be seen in Sec.~\ref{results}, where the orthonormality condition
(\ref{ortho}) of the w.fs., which are imported from {\HFODD} results,
are fulfilled to high precision. The precision of the integration
procedure can also be assessed from table~\ref{table1}, where the
density-related energies in the Skyrme terms ($E_{\rho^2}$,
$E_{\rho^{2+\alpha}}$) are rather close to that of the {\HFODD} values
even for the most coarse discretization scheme.

\section{Quantification of errors due to the FD approach to the HFB problem}
\label{results}

To assess the precision of given resolution-box-size combination, I
will take the q.p. w.f. (\ref{wf}) resulting from 3D HFB code
{\HFODD}~\cite{doba97a, doba97b, doba00, doba04, doba05, doba09,
  schunck12, schunck17}. The Skyrme force parameter SkM*~\cite{doba84}
has been used. Volume pairing is used with both proton and neutron
pairing strengths of 200 MeV fm$^{-3}$.

Defining
\begin{equation}
\label{wavefunction}
\phi_{\alpha}(\vec{r}\sigma)=
\mqty(\varphi^{(1)}_{\alpha,s=-i}(\vec{r}\sigma) \\ \varphi^{(2)}_{\alpha,s=-i}(\vec{r}\sigma)),
\end{equation}
one obtains the HFB matrix (\ref{W}) with matrix elements (M.E.)
\begin{equation}
\label{matrix}
h_{\alpha\beta} \equiv \sum_{\sigma\sigma'} \iint \dd[3]{\vec{r}}\dd[3]{\vec{r}'} 
           \phi^+_{\alpha}(\vec{r}\sigma) \mathscr{W} \phi_{\beta}(\vec{r}'\sigma').
\end{equation}
In the
present case, the w.fs. in Eq.~(\ref{wavefunction}) is a
self-consistent solution of the HFB problem in HO basis. Only states
having $s=-i$ are chosen so that the diagonal M.E. $h_{\alpha\alpha}$
are all positive, whereas the spectra for $s=+i$ are
$-h_{\alpha\alpha}$~\cite{doba04}. In this work, it has been checked
that the orthonormality condition
\begin{equation}
\label{ortho}
\sum_{\sigma}\int\dd[3]\vec{r}\phi^+_{\alpha}(\vec{r}\sigma)\phi_{\beta}(\vec{r}\sigma)=\delta_{\alpha\beta},
\end{equation}
is fulfilled up to high precision.

Ideally, if the FD representation is precise, the matrix
$h_{\alpha\beta}$ should be diagonalized as they are resulted from the
self-consistent HFB problem in HO basis. The deviation of HFB equation
(\ref{matrix}) from diagonalization reflects the error of FD
representation. The current work measures the round-off errors by
comparing the deviation of Eq. (\ref{matrix}) from a diagonalized
matrix $h^{\rm HO}_{\alpha\beta}$, which is from {\HFODD} calculation.

\begin{figure}
\centering
\includegraphics[width=0.85\columnwidth]{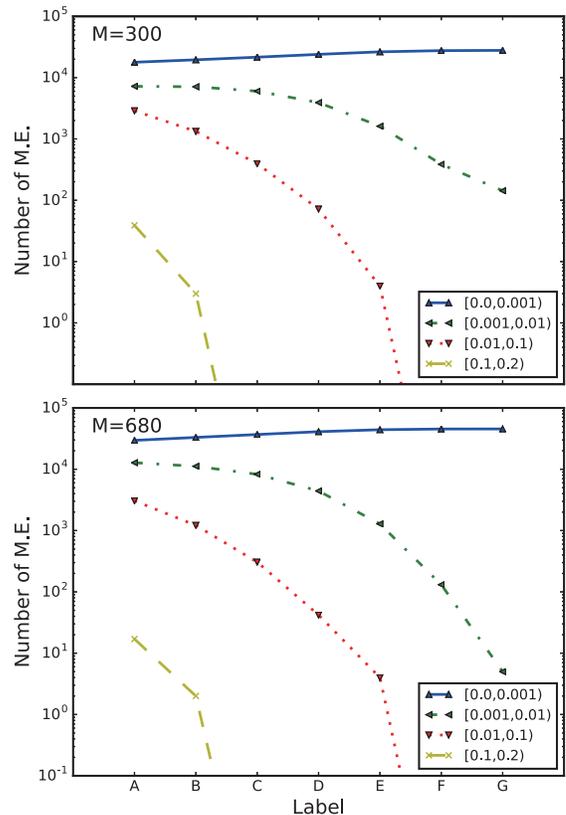}
\caption{Number of off-diagonal M.E. in the intervals of $[0, 0.001)$, 
$[0.001, 0.01)$, $[0.01, 0.1)$, and $[0.1, 0.2)$ for
discretizations `A$-$G' listed in table~\ref{table1}.
}
\label{figure2}
\end{figure}

Figure~\ref{figure2} shows the number of off-diagonal M.E. (\ref{matrix}) that
lies in the intervals [0, 0.001), [0.001, 0.01), [0.01, 0.1), and [0.1, 0.2)
for discretizing schemes listed in table~\ref{table1}. For both 300 and 680 HO
basis, there is a dominance of M.E. with values $<$0.001\,MeV, especially for
`E-G'. With decreasing grid spacing (from `A' to `G'), there is a continuous
increase of M.E. with values $<$0.001\,MeV. To better evaluate the quality that
these discretization schemes offer, one needs to have a criteria. In the
following discussions, if the number of M.E., whose values are $\ge$10\,keV, is
lower than 10$^3$, which is less than 10\% of the total number of the
off-diagonal M.E., then the corresponding discretizing scheme is considered to
be offering reasonably accurate results. With this criteria in mind, one may
find that the HFB matrix is reasonably diagonalized for grid spacings smaller
than 0.7\,fm (`C-G' schemes). From figure \ref{figure2}, it can be seen that,
for grid spacings of 0.5-0.6~fm, there are 10$^{2-3}$ M.E. with values between
1-10 keV. This means that a grid spacing of 0.5-0.6~fm provides precision on
the s.p. or q.p. levels of the order of 1-10 keV.

\begin{table}[htb]
\caption{Definition of box size and spacings used in the calculations.}
\label{table1}
\begin{ruledtabular}
\begin{tabular}{llllllll}
Label & A & B & C & D & E & F & G \\
\hline
$\Delta x,y,z$ (fm) & 1.0 & 0.9 & 0.8 & 0.7 & 0.6 & 0.5 & 0.4 \\
N$^{x,y,z}_{\rm max}$ & 14 & 14 & 16 & 18 & 22 & 26 & 32 \\
Box size (fm) & 28.0 & 25.2 & 25.6 & 25.2 & 26.4 & 26 & 25.6 \\
\end{tabular}
\end{ruledtabular}
\end{table}

\begin{figure}
\centering
\includegraphics[width=0.85\columnwidth]{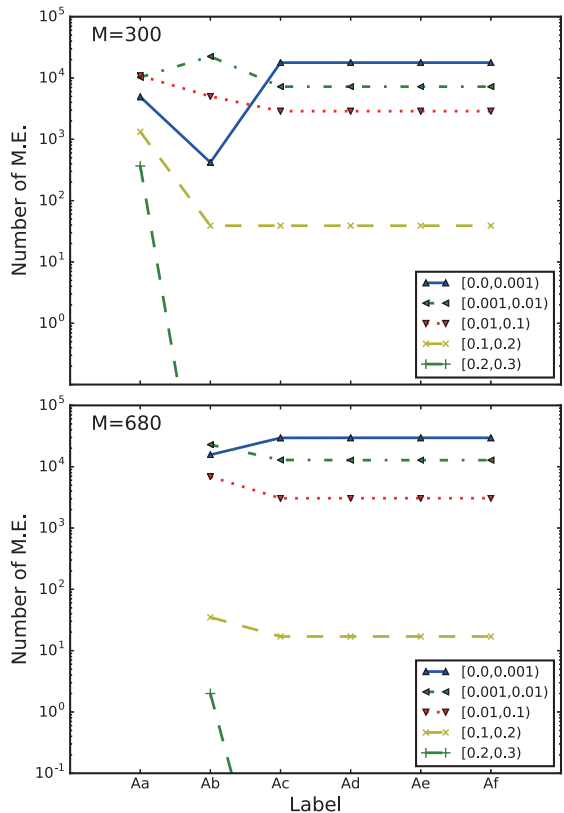}
\caption{Number of off-diagonal M.E. in the intervals 
  $[0, 0.001)$, $[0.001, 0.01)$, $[0.01, 0.1)$, $[0.1, 0.2)$, and
          $[0.2, 0.3)$. Labels `Aa' to `Af' denote results with box
            sizes ranging from 16.0 to 36.0 fm at equal distance.  }
\label{figure3}
\end{figure}

Figure~\ref{figure3} shows the number of off-diagonal M.E. for grid
spacing of 1.0\,fm as a function of box size. The distribution of
M.E. is varying for small box sizes ($\le24$ fm), and stabilizes for
box size larger than 24 fm. For a basis number of M=680 (lower panel),
the box size at which the distribution stabilizes is slightly larger
than that of M=300 (upper panel). The number of large M.E. can only be
reduced by decreasing the grid spacings. These are expected for medium
heavy nuclei from studies based on HO basis.

\begin{figure}
\centering
\includegraphics[width=0.85\columnwidth]{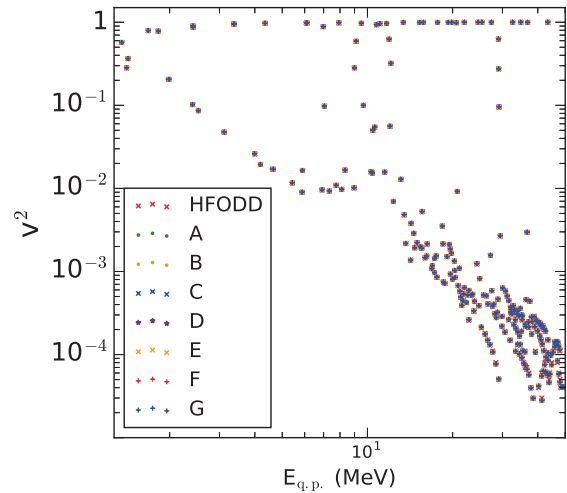}
\caption{Occupation probabilities (\ref{probability}) as a function of
  quasi-particle energies (in MeV), resulted from {\HFODD} (M=300) and
  FD (`A-F' configurations in Table.~\ref{table1}) calculations.}
\label{figure4}
\end{figure}

Figure~\ref{figure4} displays the occupation probabilities
\begin{equation}
\label{probability}
v^2_{\alpha}=\sum_{\sigma}\int\dd[3]{\vec{r}}\qty|\varphi^{(2)}_{\alpha,s=-i}(\vec{r}\sigma)|^2,
\end{equation}
against the q.p. energies obtained with {\HFODD} (M=300). For
discretization schemes `A-F', the diagonal M.E. of $h_{\alpha\alpha}$
for $\alpha=1,...,M/2$ are used. It can be seen that the q.p. spectra
for {\HFODD}, and `A-F' overlap with each other, and one could not
distinguish them from one another. In general, from scheme to scheme,
the differences of q.p. energies are mostly under 1 keV throughout the
spectra. The `F' discretization scheme spectrum is almost identical
with the one calculated with {\HFODD}. This is consistent with the
situation shown in figure~\ref{figure2}, where the error due to finite
grid spacing is $\sim0.001-0.01$\,MeV for grid spacings $\le0.6$\,fm.

Apart from the errors of s.p., or q.p. levels, which is associated
with the approximated treatment of s.p., or q.p. Hamiltonian, another
important source of error is due to the evaluation of integral
quantities such as total energies. In table~\ref{table2}, I list the
total energy of $^{110}$Mo and its decomposition into various terms,
calculations are done with FD method for discretization schemes listed
in table~\ref{table1}. The results are compared with those of {\HFODD}
calculations with M=300 which corresponds to $N$=11. For terms
containing kinetic densities in Skyrme and kinetic (\ref{kinetic})
energies, a second row has been added, where the kinetic densities,
$\tau$, is evaluated with Laplacian operator by
using~\cite{jin17}
\begin{equation}
\label{newtau}
\tau(\vec{r}) = \frac{1}{2}\Delta\rho(\vec{r}) - 
{\rm Re}\qty(\sum_{\alpha,s=\pm i}\sum_{\sigma=\pm\frac{1}{2}} 
\phi^{(2)*}_{\alpha,s}(\vec{r}\sigma)\Delta\phi^{(2)}_{\alpha,s}(\vec{r}\sigma)).
\end{equation}

\begin{table*}[htb]
\caption{Total energy of $^{110}$Mo and its decompositions into
  $E_{\rm Kin+c.m.}$ (\ref{kinetic}) and various terms in $E_{\rm
    Skyrme}$, with {\HFODD} code (M=300) and FD procedure with various
  box configurations in table~\ref{table1}. All values in the table
  are in MeV.}
\label{table2}
\begin{ruledtabular}
\begin{tabular}{lllllll}
 & C & D & E & F & G & {\HFODD}\\
\hline
\multirow{2}{*}{$E_{\rm Kin+c.m.}$}  &   2004.509   & 2007.762    & 2009.459      & 2010.240    & 2010.542  &  \\
                                     &   2010.523   & 2010.606    & 2010.639      & 2010.650    & 2010.653  & 2010.654 \\
\hline
$E_{\rm Coul}^{\rm (Dir.)}$            & 266.908      &  266.903    & 266.902       & 266.899     & 266.897   & 266.908 \\
\hline
$E_{\rm Coul}^{\rm (Exc.)}$            & $-$15.744    & $-$15.744   & $-$15.744     & $-$15.744   & $-$15.744 & $-$15.744 \\
\hline
$E_{\rm \rho^2}$                     & $-$12266.985 & $-$12266.985& $-$12266.985  & $-$12266.985&$-$12266.985& $-$12266.984 \\
\hline
\multirow{2}{*}{$E_{\rm \rho\tau}$}  & 367.613     & 368.345    & 368.729        & 368.907      & 368.976  & \\
                                     & 368.964     & 368.988    & 368.997        & 369.001      & 369.002  & 369.002 \\
\hline
$E_{\rm \rho^{2+\alpha}}$            & 8600.091    & 8600.091    & 8600.091   & 8600.091     & 8600.091 & 8600.090 \\
\hline
$E_{\rm \rho\Delta\rho}$             & 193.189     & 193.209     & 193.218       & 193.221     & 193.222  & 193.222 \\
\hline
$E_{\rm \rho\nabla J}$               & $-$73.778   & $-$73.935   & $-$74.018    & $-$74.056    & $-$74.071 & $-$74.077 \\
\hline
$E_{\rm pair}$                       & $-$3.816    & $-$3.816    & $-$3.816      & $-$3.816    & $-$3.816 &  $-$3.816 \\
\hline
\multirow{2}{*}{$E_{\rm Skyrme}$}    & $-$3179.870 & $-$3179.275& $-$3178.965   & $-$3178.822 & $-$3178.767 & \\
                                     & $-$3178.519 & $-$3178.632& $-$3178.697   & $-$3178.728 & $-$3178.741 & $-$3178.747 \\
\hline
\multirow{2}{*}{$E_{\rm Total}$}     & $-$928.014 & $-$924.172& $-$922.164   & $-$921.243 & $-$920.888 & \\
                                     & $-$920.649 & $-$920.684& $-$920.716   & $-$920.739 & $-$920.751 & $-$920.745 \\
\end{tabular}
\end{ruledtabular}
\end{table*}

As expected, the energies converge towards the {\HFODD} results with
decreasing grid spacings. It is interesting to note that, the method
used to evaluate the kinetic density in FD method, using (\ref{tau})
or (\ref{newtau}), impacts the speed of convergence. With
Eq.~(\ref{tau}), for `G' configuration, the total energy differs from
the {\HFODD} result by $\le$200\,keV. Using Eq.~(\ref{newtau}), the
energies converge to {\HFODD} result at a much larger grid spacing
(0.7\,fm) than the results given by Eq.~(\ref{tau}).

In the following discussions, a guideline will be defined to assess
these discretization schemes: if an energy calculated with certain box
discretization differs from the value calculated with {\HFODD} by
$\le$100\,keV, then the respective discretizing scheme is considered
to be reasonably accurate. It is then noticed that the terms
containing differential operations ($E_{\rm Kin+c.m.}$, $E_{\rm
 \rho\tau}$, $E_{\rm \rho\Delta\rho}$, and $E_{\rm \rho\nabla J}$)
shows lowered precision for discretization schemes `C'. In particular,
according to the values of $E_{\rm Kin+c.m.}$, one may find that `D-G'
configurations with $\Delta x,y,z\le$0.7\,fm provide much better
approximation than that of `C'. This situation contrasts with that
shown in figures \ref{figure2}, and \ref{figure3}, where the
discretization scheme `C' already provide rather good accuracy on the
q.p. levels, with majority of the M.E. being smaller than 10\,keV.


\section{Self-consistent calculations}
\label{examples}

\subsection{self-consistent HF calculations}
\label{HF}

In this section, I examine the results when self-consistency has been
achieved, by comparing results calculated in the current work with
that of {\HFODD}. As has been noted in the introduction, there are
ambiguities for calculations represented in different dimensions. This
is due to different truncation schemes. These differences make a
stringent benchmark calculation difficult. HF calculations for light
doubly magic nuclei are probably most suited to perform benchmark
calculations, where the differences in s.p. space are minimized.

I perform self-consistent HF calculations for a well defined light
nucleus $^{40}$Ca, and a heavier nucleus $^{132}$Sn. The results are
tabulated in table \ref{table3}. Although being fundamentally
different in the pairing treatment, the sources of error due to box
discretization for both methods are rather similar. This is because
neither the pairing density (\ref{pdensity}), nor pairing Hamiltonian
(\ref{phamiltonian}) contain any differential operation, which is the
main source of error of FD method.

The self-consistent HF problem is solved by updating w.fs. using an
imaginary time step method~\cite{davies80}, which provides stable
converged solution. Iteration is terminated when sum of the
dispersions of the single particle energies, $\frac{1}{A}
\sum_{\alpha} \qty[ (h^2)_{\alpha\alpha} - (h_{\alpha\alpha})^2 ] <
1.0 \times 10^{-5}$ for ten iterations. Note, that the $h$ here is the
s.p. Hamiltonian which appears in Eq.~(\ref{W}). With this criteria of
convergence, the difference of total energy, between the current
iteration compared to the previous one, is only 10$^{-8}$ of the total
energy of the current iteration. For the solution of the Coulomb
potential, I follow the procedure described in Refs.~\cite{bonche05,
  ryss15a}.

In fact, the current part of the implemented code, which solves the
nuclear self-consistent HF problem, differs from those in
Refs.~\cite{bonche05,ryss15a} mainly by the following items:
\begin{itemize}
\item[(1)] I do not assume any symmetry in coordinate space, although
  time-reversal symmetry is present.
\item[(2)] I do not use any interpolatory techniques to maintain the
  simplicity, flexibility, and short execution time of FD method.
\end{itemize}

\begin{table}[htb]
\caption{Calculated energies for $^{40}$Ca and $^{132}$Sn with box
  discretization schemes of ($\Delta x$, $x$)=(0.625, 30) fm (denoted
  with `O'), (0.5357, 30) fm (`P'). The last column lists results
  from {\HFODD} calculation with basis size of M=969. All values given
  are in MeV.}
\label{table3}
\begin{ruledtabular}
\begin{tabular}{llll}
   & O & P  & {\HFODD} \\
\hline
$^{40}$Ca  & & & \\
$E_{\rm tot}$ & $-$344.259 & $-$344.260 & $-$344.251 \\
$E_{\rm Kin.+c.m.}$  & 633.732 & 633.686 & 644.959 \\
$E_{\rm Coul}^{\rm (Dir.)}$  & 79.502 & 79.500 & 79.616 \\
$E_{\rm Coul}^{\rm (Exc.)}$  & $-$7.478 & $-$7.477 & $-$7.489 \\
$E_{\rm Skyrme}$  & $-$1050.0146 & $-$1049.969 & $-$1051.336 \\
\hline
$^{132}$Sn  & & & \\
$E_{\rm tot}$ & $-$1103.508 & $-$1103.535 & $-$1102.935 \\
$E_{\rm Kin.+c.m.}$  & 2442.970 & 2442.784 & 2444.223 \\
$E_{\rm Coul}^{\rm (Dir.)}$  & 359.857 & 359.837 & 360.169 \\
$E_{\rm Coul}^{\rm (Exc.)}$  & $-$18.806 & $-$18.805 & $-$18.820 \\
$E_{\rm Skyrme}$  & $-$3887.529 & $-$3887.351 & $-$3888.507 \\
\end{tabular}
\end{ruledtabular}
\end{table}

The results are compared with the {\HFODD} calculations with M=969 (N=18). The
box sizes and grid spacings are chosen to be close to the `E-F' configurations
in table \ref{table1}. It can be seen that the total energies of $^{40}$Ca are
rather close to the results with {\HFODD}. For the heavier system, $^{132}$Sn,
the FD results are 500 keV lower than the {\HFODD} calculation. This is
consistent with previous results shown in Ref.~\cite{pei14}. It is comforting
to see that the total energies shown in table \ref{table3} are the same, within
a few tens of keV, as the total energies in tables 4, and 5 in
Ref.~\cite{ryss15a}.

\subsection{self-consistent HFB calculations}

To solve the HFB problem in coordinate space, a standard method is the
two-basis method devised in Ref.~\cite{gall94}. The two-basis method is
advantageous in two ways. Firstly, it is possible to reduce the dimension of
the problem to the s.p. space that is below a cutoff energy, $\epsilon \le
\bar{e}_{\rm max}$. This allows for reducing by half the dimension of the HFB
matrix that is needed to be diagonalized in the standard HFB methods. In the
imaginary-time-evolution methods~\cite{gall94}, using two-basis method is even
more beneficial as one works with the s.p. states as basis and the largest
dimension of the problem is considerably lower than the standard HFB methods.
The price that one needs to pay is that, the calculation of M.E. is numerically
expensive. But the calculation of these M.E., which is shown in
Eq.~(\ref{HFB}), can be conveniently parallelized in multi-node computers.
Secondly, the two-basis method, using s.p. states as basis, avoids problems
related to the cutoff in the quasi-particle (q.p.) space~\cite{doba12}.

\begin{table}[htb]
\caption{Calculated energies for $^{110}$Mo with the HFB code
  developed in this work (box discretization schemes of ($\Delta x$,
  $x$)=(0.725, 34.8) fm). The results are compared with that of
  {\HFODD} calculations with basis size of M=969. The oscillator
  constant is 0.4975890\,fm$^{-1}$. The parameterization is SKM*, with
  $\hbar^2$/2m=20.73\,MeV\,fm$^2$. For `b', the cutoff energy on
  s.p. levels is 10\,MeV; for `a' and {\HFODD}, the s.p. energy cutoff
  is 20\,MeV; the pairing strengths are $V_0$=$-$250\,MeV\,fm$^{-3}$
  for both protons and neutrons. Proton pairing vanishes for all
  cases.}
\label{table4}
\begin{ruledtabular}
\begin{tabular}{llll}
    & a & b & {\HFODD} \\
\hline
$E_{\rm tot}$ (MeV)            & $-$921.954   & $-$922.004  & $-$921.846 \\
$E_{\rm Kin.+c.m.}$ (MeV)          & 2005.555     & 2005.360    & 2003.932 \\
$E_{\rm Coul}^{\rm (Dir.)}$ (MeV)  & 266.771      & 266.774     & 266.908\\
$E_{\rm Coul}^{\rm (Exc.)}$ (MeV)  & $-$15.740    & $-$15.740   & $-$15.751\\
$E_{\rm Skyrme}$ (MeV)        & $-$3176.694  & $-$3177.047 & $-$3175.965 \\
$E_{\rm pair}$ (MeV)        & $-$1.846     & $-$1.351    & $-$0.970 \\
$\Delta^{\rm n}$ (MeV)        & 0.469        & 0.403       & 0.336 \\
$\lambda^{\rm n}$ (MeV)        & $-$5.470     & $-$5.476    & $-$5.507 \\
$Q_{20}^{\rm total}$ (b)        & 298.80    & 298.45   &295.20   \\
$|Q_{22}^{\rm total}|$ (b)        & 75.46       & 75.55       &82.20     \\
\end{tabular}
\end{ruledtabular}
\end{table}

In this section, I summarize the key points of the two-basis procedure before
showing the results. For details, the reader is referred to Ref.~\cite{gall94}.
The method uses s.p. states of the mean-field Hamiltonian (\ref{mean-field}) as
basis to solve the HFB problem (\ref{HFB}). The procedure consists of two
steps. The outer routine proceeds as in the HF problem shown in
section~\ref{HF}. The inner routine includes the following procedures at each
imaginary step: the HFB (\ref{HFB}) M.E. are calculated in the s.p. states; the
canonical transformation is obtained afterwards; and finally the densities in
coordinate space is constructed. Note, that the Lagrangian multiplier is
updated in the inner routine.

In table~\ref{table4}, the calculated properties of triaxial nucleus $^{110}$Mo
are shown. The energies are compared with that of {\HFODD} with basis number
M=969. It can be seen that the total energies are the same up to a few hundreds
of keV. This is acceptable, if one notices that the two methods differ through
the basis used, one being in coordinate space, the other being in HO basis
({\HFODD}). The quadruple moments listed in table~\ref{table4} are defined in
Eqs.~(\ref{Q20},\ref{Q22}).

From table~\ref{table4}, one can see that although total energies are rather
close, each contribution differs. The pairing energies differ by as large as
$\sim$1\,MeV for `a', and {\HFODD}. This situation seems to be similar to that
in Ref.~\cite{schunck12}, where the two-basis method gives pairing energies
$\sim$1.0\,MeV different from the standard HFB results. Both of their
calculations are made with same HO basis.

Moreover, the spectrum for $\epsilon_n\ge0$\,MeV are not the same in the
present calculations and that of {\HFODD}. Specifically, in our calculation the
number of levels included in the HFB calculation ($\epsilon_n\le20$\,MeV) is
180, whereas in {\HFODD}, the number is 140. This is expected, since in the
current work, the w.fs. are expressed in real space, the discretization of the
continuum spectrum of neutrons is certainly different from the {\HFODD}
calculations, where the w.fs. are expanded in the HO basis.

Table~\ref{table4} column `b' shows the results with smaller s.p. energy
cutoff, $\bar{e}_{\rm max}\le10$\,MeV. It can be seen that the pairing
properties are now closer to the results of {\HFODD}. This comparison should
not be considered to be an effort to make these two methods identical, as the
two methods discretize the continuum in the q.p. spectra in two different ways.
Hence, the two HFB methods can {\it not} be identical, and the energy
differences between the two methods could not be considered to be due to a lack
of precision from either of the methods. It has to be noted that existing
realistic applications~\cite{tera96, yama01} of the two-basis method use an
energy cutoff of 5\,MeV above Fermi surface, which is smaller than the current
calculations.

\begin{figure}
\centering
\includegraphics[width=0.85\columnwidth]{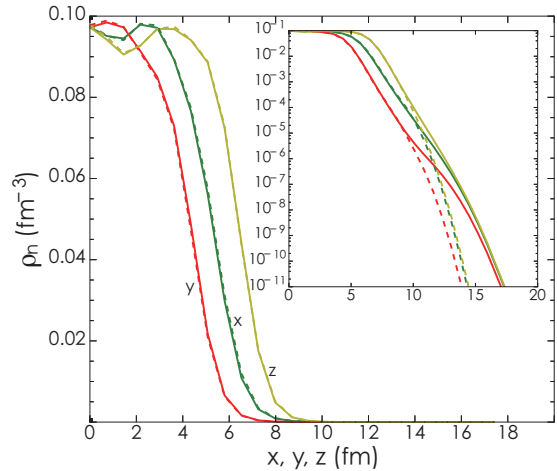}
\caption{Neutron density profiles along x-, y-, and z-axis for
  $^{110}$Mo (result `a' in table~\ref{table4}). The dashed lines are
  respective {\HFODD} results (M=680).}
\label{figure5}
\end{figure}

Figure~\ref{figure5} shows density values along x-, y-, and z-axis. It can be
seen that the HO results (dashed lines) display characteristic damping at
$r\approx10$\,fm, whereas the density values with coordinate-space code (solid
lines) start to decrease at the border of the box ($r=17.4$\,fm). Another
observation is that the lack of resolution seems to have larger influence at
the center of the nucleus. It can be expected that a box setup with more grid
density at the center part would increase precision.

\section{summary} 
\label{conclusions}

To summarize, the current work studied the precision of finite difference (FD)
approximations to the three-dimensional (3D) nuclear Hartree-Fock-Bogoliubov
(HFB) problem. To single out the error due to basis discretization, I first
performed the HFB calculations in 3D harmonic oscillator (HO) in Cartesian
coordinates. The obtained wave functions (w.fs.), together with the derivative
and Laplacian operators, were then constructed using FD method. The error was
evaluated by examining the deviation of the obtained HFB equation from a
diagonalized one, the deviation of the diagonal matrix elements (M.E.) from
those resulted from HO basis, as well as the deviation of the total energies
from the results calculated with HO basis method.

It is surprising to note that the HFB matrix resulted from FD method is rather
close to the one obtained from HO basis, with a grid spacing of 0.8\,fm.
Specifically, in the obtained HFB matrix, the number of off-diagonal M.E. that
are smaller than 0.1\,MeV dominates, whereas the diagonal M.E. have values in
the order of several to a few tens of keV. The number of non-zero M.E. decrease
rapidly with decreasing grid spacing. The error of q.p. levels is rather small:
the difference for the spectra, with varying grid spacing, is only a few keV in
general. Total energy was calculated to be only $\sim$10\,keV different from
the one resulted from HO basis at $\Delta x$=0.6\,fm. For a discretizing space
of 0.7\,fm, each contribution of the energy difference with respect to the HO
result was lower than 100\,keV. This provided acceptable precision for the FD
method in the current work. Results with larger grid spacings ($\Delta x \le
0.8$\,fm) exhibited an important loss of accuracy. For evaluation of kinetic
energy, the importance using Eq.~(\ref{newtau}) instead of Eq.~(\ref{tau}) was
emphasized.

The effect of self-consistency was studied by performing self-consistent HF
calculations for doubly magic nuclei, $^{40}$Ca, and $^{132}$Sn. The results
for grid spacing $\sim$0.6\,fm were consistent with results obtained with HO
basis method, and existing similar 3D coordinate space methods. In addition,
the self-consistent HFB calculations, in the two-basis method, for triaxially
deformed nucleus $^{110}$Mo were performed. The results were compared with
those from {\HFODD}. It was demonstrated that the spectra for the two models
(coordinate space and HO) differ for positive energy values.

\begin{acknowledgments}

Useful discussions with M. Bender, D.L. Fang, Y. Gao, M. Kortelainen, 
J.C. Pei, J. Toivanen, and C.X. Yuan are gratefully acknowledged.
I thank N. Michel, and W. Nazarewicz for their careful reading of the manuscript and useful comments.
The current work is supported by National Natural Science Foundation of China (Grant No. 11705038).
I thank the HPC Studio at Physics Department of Harbin Institute of 
Technology for computing resources allocated through INSPUR-HPC@PHY.HIT.

\end{acknowledgments}

\end{document}